\documentclass[apjl]{emulateapj} \usepackage{times}
\bibpunct{(}{)}{;}{a}{}{,} 

\newcommand{\be}{\begin{equation}}
\newcommand{\ee}{\end{equation}}

\newcommand{\remove}[1]{}

\newcommand{\V}{$V^2$ }

\begin{document}

\shortauthors{Pinte et al.}
\title{The inner radius of T Tauri disks estimated from near-infrared
  interferometry:\\ the importance of scattered light.}
\shorttitle{Inner radius of T Tauri disks estimated from NIR
  interferometry: the importance of scattered light.}

\author{Pinte, C.\altaffilmark{1}, M\'enard, F.,
  Berger, J.P., Benisty, M., Malbet, F.} 
\email{pinte@astro.ex.ac.uk}
\affil{Laboratoire d'Astrophysique de Grenoble, CNRS/Universit\'e Joseph-Fourier, UMR~5571, 
B.P. 53, F-38041 Grenoble Cedex 9, France}
\altaffiltext{1}{School of Physics, University of Exeter, Stocker Road, Exeter EX4 4QL, United Kingdom}

\begin{abstract}
  For young Herbig AeBe stars, near-infrared interferometric
  measurements have revealed a correlation between the luminosity of
  the central object and the position of the disk inner rim. This
  correlation breaks down for the cooler T Tauri stars, a fact
  often interpreted in terms of disks with larger inner radii.
  In most cases, the conversion between the observed interferometric
  visibility and the calculated disk inner radius was done with a
  crude disk emission model. Here, we examine how the use of models
  that neglect scattered light can lead to an overestimation of the
  disk sizes. To do so, synthetic disk images (and visibilities) are
  calculated with a full treatment of the radiative transfer. The
  relative contributions of thermal emission and scattered light are
  compared. We find that the latter can not be neglected for cool
  stars. For further comparison, the model visibilities are also
  converted into inner disk radii using the same simple disk models as
  found in the literature. We find that reliable inner radii can only
  be estimated for Herbig~Ae/Be stars with these models. However, they
  lead to a systematic overestimation of the disk size, by a factor of
  2 to 3, for T~Tauri stars. We suggest that including scattered light
  in the models is a simple (and sufficient) explanation of the
  current interferometric measurements of T Tauri stars.

\end{abstract}

\section{Introduction}

Near-infrared (NIR) broad-band interferometric observations of young
stellar objects trace the inner part of the warm dusty circumstellar
environment ($\lesssim 1$\,AU) and can be used to constrain its
physical properties and to characterize the size and location of the
emitting region.

The NIR sizes, derived from simple geometrical models, are found to be
consistent with the dust sublimation radius for Herbig~Ae and late Be
objects (\citealp{Monnier02,Millan-Gabet06PPV} and references
therein). Interestingly, observations of T~Tauri stars reveal lower
NIR visibilities (hereafter $V^2$). They correspond to sizes larger than the dust
sublimation radii when the same simple geometrical models are used
\citep{Akeson05,Eisner05,Eisner07}. Various explanations are proposed
to account for these surprisingly low \V (or large inner
radii): extra heating from accretion, lower dust sublimation
temperatures, small dust grains, photo-evaporation.  As noted by
\cite{Eisner07}, it is unlikely however that any of these mechanisms 
can explain by itself the large inner radii observed in all low mass
T~Tauri stars. Instead, \cite{Eisner07} favor the explanation where
the inner disk's position is controlled by the stellar magnetospheric
pressure, in this case pushing the disk outward when accretion rates
are low enough.

In this letter, we explore the possibility that the position of the
inner disk is incorrectly estimated, especially for T~Tauri stars,
because the radiative transfer (hereafter RT) schemes used are incomplete,
namely that the contribution from scattering is overlooked. Scattered
light is a spatially extended component potentially leading to lower
\V and, by way of consequence, to a biased estimation of
characteristic sizes when neglected. 
In the following, we
investigate these effects, using detailed RT modelling, by comparing
the spatial distribution of scattered light and thermal emission.

\section{Radiative transfer modelling}\label{sec:RT}

Synthetic images are computed using the 3D radiative transfer code MCFOST
\citep{Pinte06} at $2.2\,\mu$m ($K$~band), with a pixel sampling of
0.125\,mas. The temperature structure is calculated
assuming the dust is in radiative equilibrium with the local
radiation field (including both the stellar and reprocessed  contributions).
$K$~band maps
are then computed by emitting and propagating the proper amount of
stellar and disk thermal photon packets. All these packets are allowed
to scatter in the disk as often as needed.

\V are calculated assuming a Gaussian field of view of
50\,mas. This is the field of view of the Keck
interferometer. A distance of $140$\,pc is used. We use these values
for comparison with available data. We explore a range of stellar
luminosities by varying the effective temperature from $3\,000$\,K to
$10\,000$\,K, representing young stellar objects ranging from low mass
T~Tauri stars to Herbig Ae stars. The stellar radius is fixed at two
solar radii and stars are assumed to radiate as blackbodies.

\paragraph{Disk geometry.} 

We adopt the same geometry as in \cite{Pinte07},
with a surface density $\Sigma(r) =
\Sigma_0\,(r/r_0)^{-1}$, a scale height $ h(r) = 10\,
(r/r_0)^{9/8}\,$AU and $r_0 =100$ AU. The disk dust mass is set to
$10^{-4}$\,M$_\odot$ and the outer radius to 300\,AU.  The central
part of the disk being optically thick, the inner radius
($R_\mathrm{in}$) is computed
self-consistently, for each star, by moving it to the location where
the dust temperature is equal to the dust sublimation temperature
($T_\mathrm{sub} = 1\,500\,$K or $2\,000\,$K). 
At the inner edge, a radial Gaussian
profile towards the central star, with a $1/2e$ length of 0.01\,$R_\mathrm{in}$,
is used to reproduce a round-shaped inner rim, making it visible from all inclination angles.
We restrict our analysis in this Letter
to the pole-on case, without loss of generality as long as the
photosphere is directly visible by the observer.

\paragraph{Dust properties.}
We consider  homogeneous spherical
grains with sizes distributed according to the power-law $\mathrm{d}n(a)
\propto a^{-3.7}\,\mathrm{d}a$ between  $a_{\mathrm{min}} = 0.03\,\mu$m and
$a_{\mathrm{max}}$. Because the relative contributions of scattered
light and thermal emission will depend on the heating and
scattering properties of dust, we explore various dust populations, by
varying the maximum grain size $a_{\mathrm{max}} = 1\,\mu$m or 1\,mm and
the dust composition. Compact amorphous silicates grains
\citep{Dorschner95} and a porous mixture of amorphous silicates and
carbon (\citealp{Mathis89}, model~A) are considered.  Optical
properties are calculated using Mie theory
(Table\,\ref{tab:optical_prop}) and the non-isotropic nature of the
scattering is explicitly preserved in the calculations. A reference
model (Silicates grains, $a_\mathrm{max} = 1\,\mu m$, $T_\mathrm{sub}
= 2\,000\,$K) is first discussed in sections\,\ref{sec:profiles},
\ref{sec:v2}, \ref{sec:estimated_rin} and the effects of the different
dust properties are analysed in section\,\ref{sec:dust_prop}.

\begin{table}[tb]
  \begin{center}
  \caption{Optical properties at 2.2\,$\mu m$\label{tab:optical_prop}}
  \begin{tabular}{cllcc}
    \tableline
    \tableline
    & Composition &  $a_\mathrm{max}$ & albedo & g$^a$\\ 
    \tableline
    A & Silicates \citep{Dorschner95}& 1\,$\mu$m &  0.90 & 0.53\\
    B & Silicates \citep{Dorschner95}& 1\,mm &  0.80 & 0.62\\
    C & Si + C \citep{Mathis89}& 1\,$\mu$m & 0.37 & 0.58 \\
    D & Si + C \citep{Mathis89}& 1\,mm &  0.62 & 0.91\\
    \tableline
  \end{tabular}
\vspace{-4mm}
\tablenotetext{a}{Asymmetry parameter, $g = < \cos\theta>$ with
  $\theta$ the scattering angle} 
\vspace{2mm}
\end{center}
\end{table}

\section{Results}\label{sec:results}

\subsection{Brightness profiles} \label{sec:profiles}

\begin{figure}
  \centering
  \includegraphics[width=\hsize]{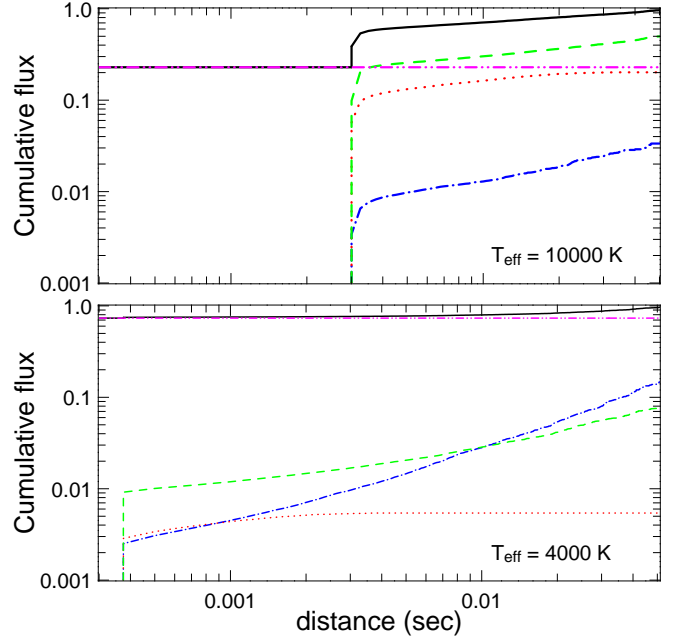}
  \caption{$K$ band integrated fluxes for a $10\,000$\,K star (upper panel) and $4\,000$\,K star (lower panel) distant of 140\,pc. Full black line :
    total flux, magenta dot-dot-dash: direct stellar light, blue
    dot-dashed: scattered stellar light, red dotted line: direct thermal
    emission, green dashed line : scattered thermal emission. The
    total flux is normalized to 1 at a distance of 50\,mas. {All contributions
    lead to a disk-to-stellar flux ratio of $\approx 3.35$ and $0.37$
    for the  $10\,000$\,K and $4\,000$\,K star respectively.}}
  \label{fig:profile}
\end{figure}

In this section we present the cumulative fluxes, integrated in a circular
aperture, as a function of distance from the star. Each
contribution, i.e., direct and scattered 
starlight and direct and scattered thermal emission from the disk, is
plotted separately for comparison. 
Two models ($T_\mathrm{eff} =
10\,000\,$K, upper panel
and $T_\mathrm{eff} = 4\,000\,$K, lower panel) are presented in
Fig.\,\ref{fig:profile}.

For the $T_\mathrm{eff} = 10\,000\,$K model, the sublimation radius is
located at 0.40\,AU from the star (3\,mas at
140\,pc). {Because the disk is warm, the thermal emission is of
the same order as the stellar emission at $K$~band. All contributions
(except the photosphere) show similar radial profiles.  Because the
disk is optically thick, the dominant contribution is the scattered
thermal emission from the disk, i.e., the photons coming from deep
inside the disk that had to scatter before reaching the surface. {\sl
Depending on radius, this contribution is as large as 2.5 times the
direct disk emission in $K$ band} because the emission volume is
larger. This contribution is often neglected.  Within a 50mas
field-of-view, direct starlight is the next contributor to the total
flux, followed by direct thermal emission from the disk}. The scattered
starlight contributes significantly less, of the order of a few \% of
the stellar flux. {It is only a small fraction of
the direct starlight because the stellar photosphere is seen directly and the disk is geometrically thin.}

For the cooler star, the sublimation radius is located closer to the
photosphere, down to $0.34$\,mas (0.048\,AU). It is only
marginally resolved by current interferometers. {In this case,
contrarily to the more massive star, scattered starlight is of the
same order as thermal emission. It can even dominate at larger
distances from the star. Such a difference comes from the fact that T
Tauri stars radiate a larger fraction of their bolometric luminosity
in the NIR and because the disk $K$ band emitting zone is smaller (i.e.,
smaller inner radius and cooler disk). Then, because the direct
thermal emission flux decreases faster with radius than scattered
light, the integrated thermal emission curve increases more slowly
than both scattered components (photospheric and disk thermal)}. A
significant effect of scattered light on \V is therefore
expected.  

\subsection{Visibilities} \label{sec:v2}
\begin{figure}
  \centering
  \includegraphics[width=\hsize]{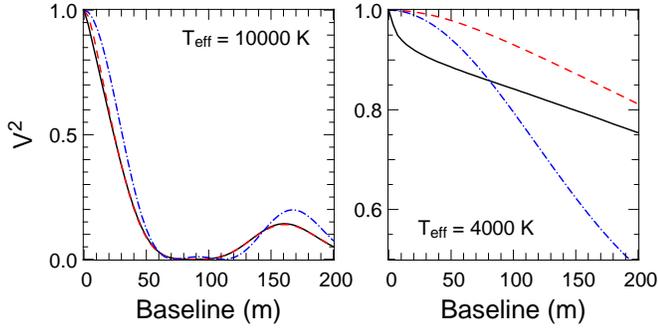}
  \caption{\V curves. Left: Herbig Ae star with
    $T_\mathrm{eff} = 10\,000$\,K. Right: T~Tauri star,
    $T_\mathrm{eff} = 4\,000$\,K. The full lines present the
    \V obtained from MCFOST images and  the dashed lines, the
    \V calculated from the direct thermal emission only (but
    with same excess). The dot-dash lines show the \V
    of the ring models which have been fitted on the \V points at 80\,m.}
  \label{fig:v2}
\end{figure}

As shown in section\,\ref{sec:profiles}, scattering can contribute
significantly to the circumstellar flux, even at small scales.  To
evaluate the incidence of neglecting scattering in the
interpretation of \V, we compare two \V curves
in Fig.\,\ref{fig:v2}: i) direct Fourier transform of the images
obtained through the MCFOST calculations (solid line), ii)
Fourier transform of the image without any scattered light
and where the thermal emission is scaled to the
total infrared excess to compensate.
This is as if all the total disk flux were emitted following the
emission profile of the direct disk thermal emission (dashed line). It ensures that the disk-to-star flux ratio remains equal in
both cases.

A comparison of both curves shows that \V are similar at
long baselines for the ${T_\mathrm{eff} = }10\,000$\,K star
(Fig.\,\ref{fig:v2}, left panel). This is an indication that most of
the emitting structure is in a small zone, at or close to the inner
rim. However, \V become different at baselines
smaller than $\sim50\,$m. Differences are of the order of a few percents for
baselines shorter than $20\,$m where the large scale structure is
completely resolved and contributes as uncoherent light, hence
reducing the $V^2$.

\V computed by taking into account the spatial 
extension of scattered light are smaller than \V computed
assuming that all the disk excess comes from direct thermal
emission. This is particularly striking for $T_\mathrm{eff} =
4\,000$\,K (Fig.\,\ref{fig:v2}, right panel).  In this case, the
discrepancy between both \V curves is dominated by the large
scale structure of both photospheric and thermal scattered light.
This large scale structure appears in the fast \V drop of
about $0.05$ occurring on baselines shorter than $\approx
20$\,m. Since the object is less resolved because the inner disk rim
is closer in, and because the contribution of large scale structures
(scattered light) to the overall excess budget is larger than in the
previous case, the discrepancy between the two curves is permanent
throughout the whole baseline range. {\em Therefore, the
interpretation of \V measurements attributing all the excess
emission to pure thermal emission is incorrect for cool stars and
previous results must be taken with care}. In particular,
since these measurements are often made with a precision of $\approx
5\,\%$ they can lead to significant errors which are discussed in the
following section.

\subsection{Estimation of inner radii from NIR visibilities} \label{sec:estimated_rin}

The decrease of \V at short baselines due to the larger
spatial extent of scattered light can result in an overestimation of
the inner radius of the disk if interpreted in the wrong context.  To
quantify this effect, we fit the \V calculated above with a
simple geometrical model composed of an unresolved point source
surrounded by a thin ring, as done, e.g.,  by \cite{Akeson05} and
\cite{Eisner05, Eisner07}. We adopt a Gaussian brightness profile in the
radial direction\footnote{The exact profile of the ring does not
matter for our study, as long as the width of the ring is small
compared to its radius.}.  The contribution (flux) of the ring is set
to the total contribution (excess) from the disk (i.e., thermal
emission + scattered light). As the models have been calculated for
pole-on disks, we restrict the study to circular rings.  Fitting is
performed on \emph{one} \V point at a baseline of 80\,m (the
average projected baseline in the observations of \citealp{Akeson05})
and the resulting \V curves are shown in Fig.\,\ref{fig:v2}
(dot-dash lines).

For the $T_\mathrm{eff} = 10\,000\,$K star, this curve remains close
to the \V curve calculated with full RT,
indicating that the derived radius (0.42\,AU) is a good estimation of
the actual inner radius (0.40\,AU). However, at short baselines the
shape of the curves differ significantly, with differences larger than
10\,\% for a 20\,m baseline. {\em Therefore, for warm stars, results
from short baseline measurements should be taken with caution since
they can be biased by extended emission (scattering).} Interestingly,
such a decrease of \V at short baselines have been observed
for a few objects with IOTA
\citep{Monnier06}\footnote{These observations were performed 
in the $H$~band but the effect of scattered light is similar.}.
{Similarly, results from longer baselines and more distant objects
(\emph{i.e.} sampling the same spatial scales as described here)
should also be interpreted with care.}

For the cooler $T_\mathrm{eff} = 4\,000\,$K star the situation is
worse. {The \V curve fitted with a ring is no longer a
good approximation of the true disk \V curve because of very
different behavior at short baselines. The inferred ring radius
(0.12\,AU) is in that case a large overestimation of the actual inner
radius of the disk (0.048\,AU).}

\begin{figure}[tb]
  \centering \includegraphics[width=\hsize]{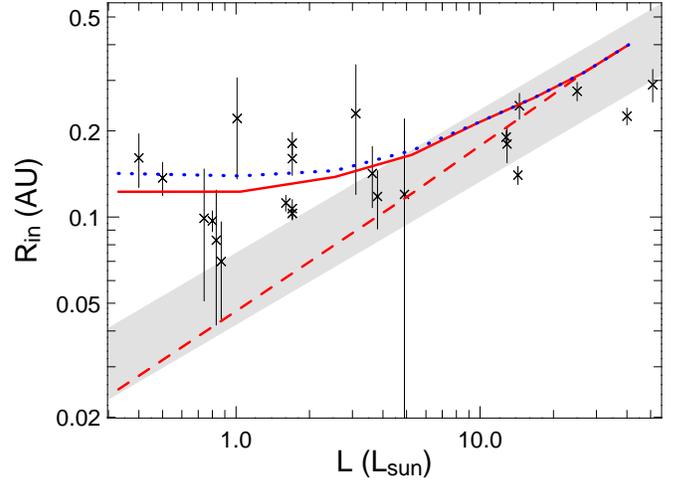}
  \caption{Comparison of the radius derived with a ring model (full
  line, field of view of 50\,mas) with the (true) physical inner
  radius of the disk (dashed line) as the function of the stellar
  luminosity. The dotted line shows the radius derived 
  assuming a field of view of 1''. Crosses correspond to
  observational data points (see Table\,\ref{tab:Rin_obs}). We omit
  error bars on luminosity to avoid crowding. The shaded area shows
  the location of the sublimation radii calculated as in
  \cite{Eisner07}, for $T_\mathrm{sub}$ between 1\,500 and
  2\,000\,K. This represents the location of the size-luminosity
  diagram where most of the Herbig Ae stars lie. Note that the
  slope of sublimation radius is larger in our
  modelling because small grains ($a_\mathrm{max}=1\,\mu$m) are used.\label{fig:KI_comparison}}
\end{figure}

Fig.\,\ref{fig:KI_comparison} shows the derived ring size as a
function of the stellar luminosity (full line), compared to the true
sublimation radius (dashed line). When neglecting scattered light,
{\sl the ring size is always an overestimation of the actual inner
disk radius} and this overestimation increases as the temperature of
the central object decreases.

Radii estimated with ring models fitted on Keck interferometer
observations are over-plotted on Fig.\,\ref{fig:KI_comparison}. The
selected sources are CTTS and Herbig stars at a distance close to
140\,pc and have been observed in $K$ band (Table\,\ref{tab:Rin_obs}).
Interestingly, the location of the fitted inner radii mimics very well
the distribution of data points. Therefore, the contribution of
scattered light appears as a simple and sufficient explanation of the
observed trend of lower \V seen in late type PMS stars given
the current error bars on the measurements.

Not all data in the literature were obtained with the same instrument
however. To estimate the effect of the interferometer's field of view, we
plot on Fig.\,\ref{fig:KI_comparison} the
derived ring size for a field of view of 1'', valid for the Palomar
Testbed Interferometer for instance.  As expected, the impact of
scattered light is more pronounced, with a derived radius of 0.14\,AU
for $T_\mathrm{eff} = 4\,000\,$K and is increasing with decreasing
stellar temperature. However, because the brightness of scattered
light decreases rapidly with distance from the star, the effect
remains limited and significantly smaller than the one resulting from
the contribution inside the inner 50\,mas.

\subsection{Effect of dust properties}\label{sec:dust_prop}
\begin{figure}[tb]
  \centering
  \includegraphics[width=\hsize]{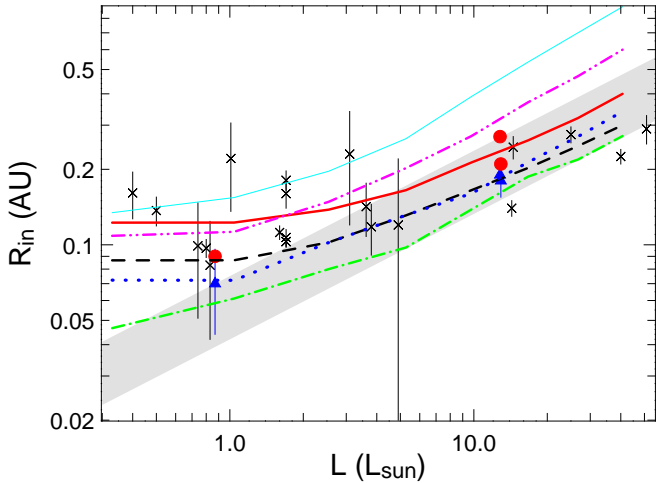}
  \caption{Measured inner radii as a function of the stellar
    luminosity for different dust populations: dust model A
    (red full line), B (black dashed line), C (blue dotted
    line), D (green dot-dashed line) with
    T$_\mathrm{sub} = 2\,000\,$K and dust models A (cyan thin full
    line) and B (pink dot-dot-dashed line) with
    T$_\mathrm{sub} = 1\,500\,$K . Crosses are observations from
    literature, blue triangle and red circles are the sizes derived
    by \cite{Akeson05b} with ring models and RT models
    respectively.}\label{fig:KI_comparison2}
\end{figure}

In order to test the dependence of the findings presented in
section\,\ref{sec:estimated_rin} on dust properties,
Fig.\,\ref{fig:KI_comparison2} shows the derived radii for several
dust populations. The general behavior remains unchanged and the
effect of scattered light on \V is significant for all dust
populations we treated.

\begin{table}[tb]
  \begin{center}
  \caption{Selected sources}
  \label{tab:Rin_obs}
  \begin{tabular}{ccccc}
    \tableline
    \tableline
    Name & distance & L (L$_\odot$) & $R_\mathrm{ring} $ (AU) & Reference\\
    \tableline
CI Tau	 & 140\,pc &   0.8  &   0.097 $\pm$  0.008   &    {1}\\
DK Tau A & 140\,pc &   1.7  &   0.103 $\pm$  0.005   &    {1}\\
DK Tau A & 140\,pc &   1.7  &   0.107 $\pm$  0.008   &    {1}\\
DK Tau B & 140\,pc &   0.5  &   0.137 $\pm$  0.018   &    {1}\\
RW Aur A & 140\,pc &   1.7  &   0.103 $\pm$  0.005   &    {1}\\
RW Aur A & 140\,pc &   1.7  &   0.181 $\pm$  0.016   &    {1}\\
RW Aur B & 140\,pc &   0.4  &   0.161 $\pm$  0.034   &    {1}\\
V1002 Sco& 160\,pc &   3.8  &   0.118 $\pm$  0.027   &    {1}\\
AS 206	 & 160\,pc &   1.6  &   0.112 $\pm$  0.007   &    {1}\\
BP Tau	 & 140\,pc &   0.83 &   0.083 $\pm$  0.041   &    {2}\\
DG Tau   & 140\,pc &   3.62 &   0.142 $\pm$  0.034   &    {2}\\
GM Aur	 & 140\,pc &   1.01 &   0.221 $\pm$  0.085   &    {2}\\
LkCa15	 & 140\,pc &   0.74 &   0.099 $\pm$  0.048   &    {2}\\
RW Aur	 & 140\,pc &   1.7  &   0.160 $\pm$  0.020   &    {2}\\
AS 207A	 & 160\,pc &   3.1  &   0.23  $\pm$  0.11    &    {3}\\
V2508 Oph& 160\,pc &   4.9  &   0.12  $\pm$  0.10    &    {3}\\
SA 205A	 & 160\,pc &   14.3 &   0.14  $\pm$  0.01    &    {3}\\
MWC 758  & 150\,pc &   25 $\pm$ 4  &   0.275  $\pm$  0.02    &    {4}\\
HD 144432& 145\,pc &   14.5 $\pm$ 4&   0.245  $\pm$  0.025    &    {4}\\
HD 150193& 150\,pc &   51  $\pm$ 17 &   0.29  $\pm$  0.075    &    {4}\\
MWC 275	 & 122\,pc &   40  $\pm$ 8 &   0.225  $\pm$  0.015    &    {4}\\
    \tableline
\vspace{-5mm}
\tablenotetext{}{{\sc References:} 1, \cite{Eisner07} ; 2, \cite{Akeson05} ;
   3, \cite{Eisner05} ; 4, \cite{Monnier05}.} 
 \end{tabular}
  \end{center}
\end{table}

{However, finer differences may be linked to the dust 
properties}. First, the position of the inner rim depends on the dust
properties. The effect is to shift the curves vertically on
Fig.\,\ref{fig:KI_comparison2}. Smaller grains and silicates are more
efficiently heated and the corresponding sublimation radius is
increased. Similarly, different materials with smaller sublimation
temperatures also result in larger inner radii. Second, because of
different scattering properties (albedo and phase function), the
fraction of scattered light is larger for silicates grains, and so is
the ratio of the measured radius over the actual inner radius.

Silicates dust, models A and B, provides a good match to the data,
although we have not explored all possibilities. Intermediate dust
properties should also agree with the data points. Moreover, with the
density structure we adopted, resulting in very opaque central
regions, a sublimation temperature of $1\,500\,$K yields inner radii
that are too large to reproduce the observed trend at high luminosity
(shaded region).

Interestingly, detailed modelling of PTI interferometric observations
has been performed by \cite{Akeson05b}, including light
scattering. The authors concluded that RT models have similar radii
than geometrical models. They use dust properties with an albedo of
0.5, $g=0.6$ (intermediate to our dust properties C and D),
$T_\mathrm{sub} = 1\,600\,$K and their less luminous source (DR Tau)
has a luminosity of 0.87\,L$_\odot$. With similar parameters, we do
indeed find that the overestimation of the inner radius remains
limited (about 30-40$\,\%$), in agreement with the results of
\cite{Akeson05b} who found a difference of $\approx 30\,\%$ for DR Tau
between RT and ring models (see triangles and circles in
Fig.\,\ref{fig:KI_comparison2}). {Unfortunately, this result was
used as a validation for all sources. Should they have used different
dust properties or considered lower luminosity sources, the
discrepancy between the two methods would have been more striking.
Because the effect of scattered light is systematic (always a decrease
of $V^2$) and always larger than 30\,\% with respect to ring
models, i.e. larger than observational error bars, we suggest that
scattered light should always be included in the analysis of NIR
$V^2$: the cooler the object the more needed scattered light
is}.

\section{Conclusions}\label{sec:discussion}

We have investigated the effect of a complete treatment of RT in
the interpretation of $V^2$. We find scattered
light is an important contribution to the disk excess and dominates for
low luminosity/cool objects. In particular, we have shown this
contribution to lead to a systematic decrease of the observed
$V^2$. \emph{Interpreted in the wrong context, the lower \V
can be mistaken with disks of larger inner radii.} For more luminous
objects ($>20$\,L$_\odot$), the effect becomes significant only at
short baselines ($ < 20$\,m) but care must also be taken.

Depending on the adopted dust properties, part, if not all of the
observed trend of estimated large radii of late type objects can be
explained by the contribution of scattered light, without requiring
any other mechanism than dust sublimation to truncate the inner disk.

However, there are several issues to consider to estimate
quantitatively the exact position of the inner boundary of a given
disk. The dust properties have an impact on the actual location of the
disk sublimation radius and on the fraction of scattered light sent to
the observer. Since the dust properties are not known accurately, the
position of the sublimation radius remains slightly uncertain.
Additional effects, like viscous heating as part of the accretion
process for example, might also lead to a different disk temperature
profile, hence to a different emission profile. A more detailed case
by case modelling effort together with additional constraints from
other observations, such as mid-infrared spectroscopy or
spectro-interferometry, should be of great help to overcome these
degeneracies.




\bibliographystyle{aa}
\bibliography{biblio.bib}

\end{document}